\documentclass[10pt,journal,compsoc]{IEEEtran}
\ifCLASSOPTIONcompsoc
  \usepackage[nocompress]{cite}
\else
  \usepackage{cite}
\fi
\ifCLASSINFOpdf
\else
\fi
\usepackage{amsmath}
\usepackage{algorithmic}
\usepackage{url}
\usepackage{subfiles} % Best loaded last in the preamble
\usepackage{multirow}
\usepackage[ruled,linesnumbered]{algorithm2e}
\usepackage{listings}
\usepackage{tcolorbox}
\usepackage{subcaption}

\usepackage{subfloat}
\usepackage{color,xcolor,colortbl}
\usepackage{xspace}
\usepackage{enumitem}
\usepackage{graphicx}
\usepackage{xr}
\usepackage{url}
\usepackage{booktabs}
\usepackage{diagbox}
\usepackage{array}
\usepackage{tcolorbox}

\newcommand{\tool}{LIVABLE\xspace}

\newcommand{\et}{\textit{et} \textit{al.}\xspace}

\newcommand{\http}{\url{https://github.com/LIVABLE01/LIVABLE}}

\begin{document}

%\title{Vulnerability Detection with Heterogeneous Graph}

\title{\tool: Exploring Long-Tailed Classification of Software Vulnerability Types}


%\author{Xin-Cheng~Wen, Cuiyun Gao, Jiaxin Ye, Yichen Li, Zhihong Tian, Yan Jia, and Xuan Wang
\author{\IEEEauthorblockN{Xin-Cheng Wen\IEEEauthorrefmark{2},
Cuiyun Gao\IEEEauthorrefmark{1}\thanks{* corresponding author.}\IEEEauthorrefmark{2}\IEEEauthorrefmark{3},
Feng Luo\IEEEauthorrefmark{2},
Haoyu Wang\IEEEauthorrefmark{4},
Ge Li\IEEEauthorrefmark{5},
and Qing Liao\IEEEauthorrefmark{2}
} \\
\IEEEauthorblockA{\IEEEauthorrefmark{2}}Harbin Institute of Technology, Shenzhen, China\\
\IEEEauthorrefmark{3} Peng Cheng Laboratory, Shenzhen, China\\
\IEEEauthorrefmark{4} Huazhong University of Science and Technology, Wuhan, China\\
\IEEEauthorrefmark{5} Peking University, Beijing, China\\ 

\IEEEauthorblockA{xiamenwxc@foxmail.com, \{gaocuiyun, liaoqing\}@hit.edu.cn, 190110308@stu.hit.edu.cn, 
haoyuwang@hust.edu.c, lige@pku.edu.cn,}}
\IEEEtitleabstractindextext{%
\begin{abstract}

% Software vulnerability detection is a fundamental problem in software security. 
Prior studies generally focus on software vulnerability detection and have demonstrated the effectiveness of Graph Neural Network (GNN)-based approaches for the task. Considering the various types of software vulnerabilities and the associated different degrees of severity, it is also beneficial to determine the type of each vulnerable code for developers. In this paper, we observe that the distribution of vulnerability type is long-tailed in practice, where a small portion of classes have massive samples (i.e., head classes) but the others
contain only a few samples (i.e., tail classes). Directly adopting previous vulnerability detection approaches tends to result in poor detection performance, mainly due to two reasons.
 %First, it is difficult to effectively learn the representation of each vulnerability type due to the serious imbalance between the numbers of samples for different types and the over-smoothing issue of GNNs.
 First, it is difficult to effectively learn the vulnerability representation due to
 % the \yun{complicated structural information in vulnerable
 % deep levels in \yun{structural} graphs generated from source code and 
the over-smoothing issue of GNNs.
Second, %\yun{tail} 
vulnerability types {in tails} are hard to be predicted due to the extremely few associated samples.

To alleviate these issues, we propose a \textbf{L}ong-ta\textbf{I}led software \textbf{V}ulner\textbf{AB}i\textbf{L}ity typ\textbf{E} classification approach, called \textbf{\tool}. \tool mainly consists of two modules, including (1) vulnerability representation learning module, which improves the propagation steps in GNN to distinguish node representations by a differentiated propagation method.
% by preserving the initial node information
%GNN by introducing
% \wxc{a differentiated propagation method for further capturing vulnerable nodes in the graph}. 
A sequence-to-sequence model is also involved to enhance the vulnerability representations. (2) adaptive re-weighting module, which adjusts the learning weights for different types according to the training epochs and numbers of associated samples by a novel training loss. We verify the effectiveness of \tool in both type classification and vulnerability detection tasks. For vulnerability type classification, the experiments on the Fan \et dataset show that \tool outperforms the state-of-the-art methods by 24.18\% in terms of the accuracy metric, and also improves the performance in predicting tail classes by 7.7\%. To evaluate the efficacy of the vulnerability representation learning module in \tool, we further compare it with the recent vulnerability detection approaches on three benchmark datasets, which shows that the proposed representation learning module improves the best baselines by 4.03\% on average in terms of accuracy.

\end{abstract}

}

\maketitle

\IEEEdisplaynontitleabstractindextext

% \fi
\IEEEpeerreviewmaketitle

\IEEEraisesectionheading{\section{Introduction}\label{sec:introduction}}

Software vulnerabilities are important and common security threats in software systems. These vulnerabilities can be easily exploited by attackers and have the potential to cause irreparable damage to software systems~\cite{Google}. For example, buffer overflow issues~\cite{CWEID119} allow attackers to exploit vulnerabilities to tamper with memory data or gain control of the system. Vulnerabilities are inevitable for many reasons, e.g. the complexity of software and the steady growth in the size of the Internet~\cite{microsoft}. Vulnerability detection has received intensive attention in the software community recently.
% , and human negligence. 
% In recent years, code vulnerability detection has received increasing attention.
%Code vulnerabilities are significant and common security threats in software systems. Thus, code vulnerability detection is a fundamental task in software system security.  The vulnerabilities are easily exploitable by attackers, potentially resulting in irreparable harm to software systems. For instance, carelessness in the handling of memory can lead to buffer overflow issues, allowing an attacker to exploit the vulnerability to tamper with memory data or gain control of the system, and so on for other purposes.

% 以往VD方法 graph的作用
% The traditional approach has been to use manually-defined patterns to detect vulnerabilities~\cite{DBLP:journals/tse/SuiYX14/,DBLP:conf/ccs/YamaguchiWGR13/,DBLP:conf/sp/YamaguchiMGR15/,DBLP:conf/sp/BackesKR09/,DBLP:conf/icse/YanSCX18/,DBLP:conf/aplas/SuiYXY11/}, but such methods rely on a large number of defined features by human experts. 
% In recent years, 
With the development of deep learning (DL) techniques, various DL-based vulnerability detection methods 
% based on deep-learning (DL) techniques 
have been proposed~\cite{DBLP:conf/ndss/LiZXO0WDZ18/, DBLP:conf/icmla/RussellKHLHOEM18/, DBLP:journals/tdsc/0027ZX0ZC22/, DBLP:conf/nips/ZhouLSD019/, DBLP:journals/tse/ChakrabortyKDR22/,DBLP:conf/sigsoft/Li0N21/}, which aim to {detect}
% report 
whether a code function
% code snippet 
is vulnerable or not.
%obtain a better code representation for vulnerability prediction \yun{[???]}. They generally process the source code as token sequences or code structure graphs. For example,
For example, SyseVR~\cite{DBLP:journals/tdsc/0027ZX0ZC22/} combines multiple sequence features to generate code slices and uses a bidirectional Recursive Neural Network (RNN)~\cite{rnn} to detect vulnerabilities. Recent studies~\cite{DBLP:conf/nips/ZhouLSD019/, DBLP:journals/tse/ChakrabortyKDR22/, DBLP:conf/sigsoft/Li0N21/} have demonstrated that
% the 
Graph Neural Networks (GNNs)~\cite{GNN} are effective in vulnerability detection by
% play a vital role in 
capturing the structural information of source code.
% from the code snippets, which have achieved state-of-the-art performance in vulnerability detection. 
For example, Devign~\cite{DBLP:conf/nips/ZhouLSD019/} applies the Gated Graph Neural Network (GGNN)~\cite{ggnn} to learn the representation of code structure graph for vulnerability detection, where the code structure graph is the combination of {Abstract Syntax Tree (AST), Control Flow Graph (CFG), Data Flow Graph (DFG) and Natural Code Sequence (NCS).}
% leveraging the code structure graph\yun{, which combines [xxx]}
% , and directly adopts a Gated Graph Neural Network (GGNN)~\cite{ggnn} to learn the representations of graphs.
% Compared with the token sequences model, these methods can represent structural information from the source code.

\begin{figure}[t]
	\centering
    \includegraphics[width=0.48\textwidth]{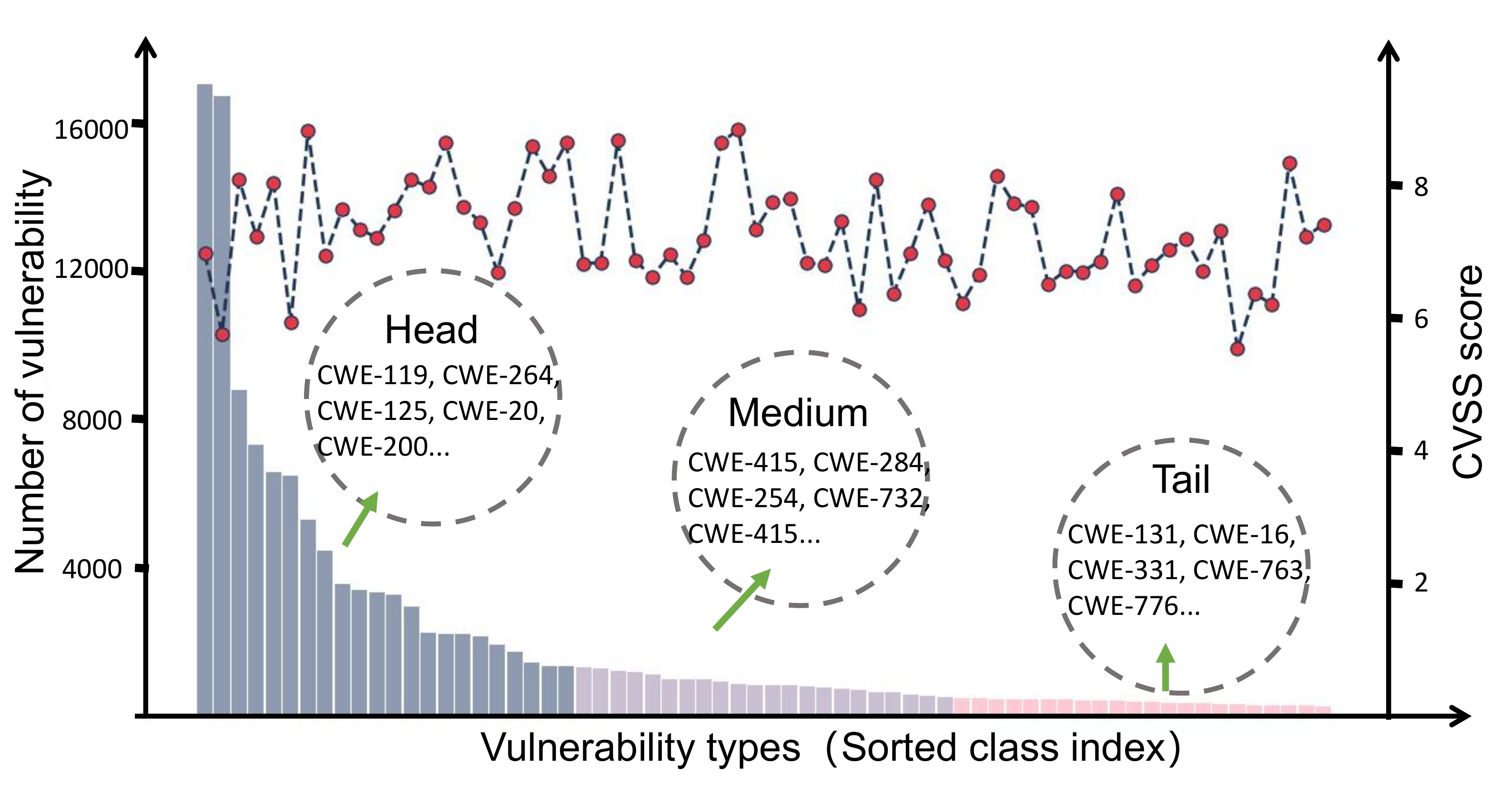}
	\caption{The label distribution (bar chart) and the corresponding CVSS score (line chart) of National Vulnerability Database (NVD) data~\cite{nvd} during the period of 2012 to 2022.% with 325 classes.
 The head, medium, and tail classes respectively %account for 85\%, 10\%, and 5\% of the total sample, which 
    contain 27, 45 and 253 classes in NVD respectively.}
 %The head (>1000 samples per class), medium (200-1000 samples per class), and tail (<200 samples per class) classes contain
 % have 35, 42 and 248
 %27, 45 and 253 classes in NVD, respectively.}
	\label{nvd_bar}
\end{figure}

% Recently, deep learning (DL) has demonstrated effectiveness in a range of tasks. In light of that, DL-based vulnerability detection methods are increasingly becoming prevalent. Gradually, traditional static approaches~\cite{} are replaced by intelligent models for the reason that: (1) DL-based methods are capable of detecting a greater variety of vulnerabilities. (2) The intelligent models have the capability to learn vulnerability patterns automatically to counter the emerging new type of patterns. Russell \et~\cite{} utilize Convolutional Neural Networks (CNNs) to build the model. Devign~\cite{} leverages Gated Graph Neural Network (GGNN) to classify the graphs transferred from the source code.

%Each of these methods only . 
Despite that the previous studies
% on vulnerability detection~
\cite{DBLP:conf/nips/ZhouLSD019/,DBLP:journals/tse/ChakrabortyKDR22/,DBLP:conf/sigsoft/Li0N21/} can inform developers of the vulnerable functions {in source code}, the developers may still feel difficult to fix the vulnerabilities. Considering the various types of vulnerabilities and the associated different degrees of severity~\cite{DBLP:conf/icinfa/ShuaiLLZT13/vc1, DBLP:conf/bwcca/NaKK16/vc2}, predicting the vulnerability types can assist developers in arranging the maintenance priority and localizing the vulnerable cause. Recently, Zou \et~\cite{DBLP:journals/tdsc/ZouWXLJ21/tdsc} %use callee relation between functions for constructing code gadget, which 
leverage function call information to conduct multi-class vulnerability detection, which requires project-level information and can hardly be applied to predict the vulnerability type of source code in function-level.
% adopts the deep learning-based system for vulnerability type classification~\cite{cheng2022bug}. But they are limited by the program information and the relationship between functions to build the code gadget. 
% However, 
To our best knowledge, no prior research has explored the type classification problem for vulnerable functions.
% in function-level. 
% \haoyu{maybe we should be cautious for some sentences like ``to our best knowledge, no prior research has explored the type classification for vulnerable code snippets'', as some previous studies have claimed that they have considered the vulnerability classification. For example,  https://arxiv.org/abs/2001.02334}

% Specifically, 
For facilitating analysis, we first study
% analyze 
the collected National Vulnerability Database (NVD)~\cite{nvd} data during the period of 2012 to 2022, as illustrated in Figure~\ref{nvd_bar}. We observe that the sample size of different vulnerability types generally presents a long-tailed distribution. Specifically, only a small portion of classes have massive samples (i.e., head classes) while the others contain extremely few samples (i.e., tail classes). Although the vulnerable code in tails is limited in number, the degree of severity would be high. According to \emph{Common Vulnerability Scoring System (CVSS)}~\cite{CVSSSIG}, which characterizes the degree of severity into a score of 1-10, the average CVSS score \footnote{In this paper, we use the \emph{Common Vulnerability Scoring System V3 Score~\cite{CVSS3}}.
% as a criterion.
} of all the classes in  tails is 7.01, indicating a high-level severity~\cite{CVSS3level}. Some of the most threatening vulnerability types in tails such as CWE-507 and CWE-912 are with a CVSS score of 9.8. For example, the CWE-507 (Trojan Horse)~\cite{CWEID507} is a virus that is added to supply chain software by malware, leading to serious security risks. Therefore, effectively classifying the vulnerable types including the tail classes is important for software security.

% threatening vulnerability types in the NVD are CWE-507, CWE-912, CWE-228, CWE-472, CWE-187 and CWE-1059, which have a \emph{CVSS v3} score of 9.8 and are all in the tail class. For example, the CWE-507 (Trojan Horse)~\cite{CWEID507} is a virus that can be added to supply chain software by malware, which can lead to serious security risks.
%\wxc{Recently, many researchers have focused on the potential of deep neural networks of vulnerability type classification~\cite{}. However, all of these works have major limitations in learning vulnerability representation to characterize vulnerabilities of high diversity and coplexity for the long-tailed scenario. }
% Although 
Directly applying the existing vulnerable detection methods~\cite{DBLP:conf/nips/ZhouLSD019/, DBLP:journals/tse/ChakrabortyKDR22/} is one possible solution. However,
% possible, 
they tend to fail in the long-tailed scenario, because: (1) They are difficult to effectively learn the vulnerability representation due to the over-smoothing issue~\cite{lukovnikov2021improving} of GNNs. Prior studies~\cite{bottleneck} show that the performance of GNNs will degrade as the number of layers increases, leading to similar representations for different nodes. However, 
% \wxc{some of the vulnerability types have similar structures and different dangers.} 
few layers in GNNs will make the model hard to capture the structural information of vulnerable code, 
% \wxc{in different types of vulnerabilities}, 
due to the deep levels in the code structure graph~\cite{graphcodebert, DBLP:conf/icse/ZhangWZ0WL19/ASTNN}. (2) It is hard to predict the vulnerability types in tails which are associated with extremely few samples. The serious imbalance between the numbers of different vulnerability types renders the models biased towards head classes, leading to poor representations of tail classes. The models are prone to producing head classes while ignoring the tails.

To mitigate the above challenges, we propose a \textbf{L}ong-ta\textbf{I}led software \textbf{V}ulner\textbf{AB}i\textbf{L}ity typ\textbf{E} classification approach, called \tool. \tool mainly consists of two modules: (1)
% We propose 
\textbf{a vulnerability representation learning module}, which involves a differentiated propagation-based GNN for alleviating the over-smoothing problem. For further enhancing the vulnerability representation learning, a sequence-to-sequence model is also involved to capture the semantic information.
% And we involve  to capture semantics information for effectively learning the vulnerability representation. 
(2)
% We also propose 
\textbf{an adaptive re-weighting module}, which adaptively updates the learning weights according to the vulnerable types and training epochs. The module is designed to well learn the representations of tail classes based on the limited samples. 

% to effectively learn the tail class vulnerabilities, which learns weights for different types according to the training epochs and numbers of the associated sample.
%Thus in this paper, we propose: (1) We leverage a novel loss function based on the long-tailed problem. Different strategies were employed for \lf{easily classified samples and hard to classify samples.} (2) Combining token-based methods and enhanced graph-based methods, we leverage a new backbone that is capable of obtaining a more semantical and structural code representation. \lf{Through this backbone, we significantly improve the shortcomings of the prior methods in feature extraction.}
%We split the existing label distribution into the head, medium (sample < 200), and tail (sample < 50) classes and evaluate the accuracy of the different classes separately. 

% We have conducted experiments 
The experimental evaluation is performed on both vulnerability type classification and vulnerability detection tasks. For vulnerability type classification, we prepare the evaluation dataset by extracting the vulnerable
% vulnerability 
samples from Fan \et~\cite{DBLP:conf/msr/FanL0N20/}.
% to evaluate the performance of existing techniques in the long-tailed data distribution vulnerability type classification setting. 
% Experiments 
The results demonstrate that the \tool outperforms the state-of-the-art methods by 24.18\% in terms of the accuracy metric, by improving
% and also improves 
the performance in predicting medium and tail class by 14.7\% and 7.7\%, respectively.  To evaluate the efficacy of the vulnerability representation learning module in LIVABLE, we further compare with the recent vulnerability detection approaches on FFMPeg+Qemu~\cite{DBLP:conf/nips/ZhouLSD019/}, Reveal~\cite{DBLP:journals/tse/ChakrabortyKDR22/}, and Fan \et~\cite{DBLP:conf/msr/FanL0N20/} datasets, which shows that the proposed representation learning module improves the best-performing baselines by 4.03\% on average in terms of accuracy.
%First, we extract a new dataset from a large-scale popular dataset Fan \et to evaluate the performance of existing techniques in the long-tailed data distribution vulnerability type classification setting. Based on the number of different vulnerability types, we split the existing label distribution into the head, medium (sample < 200), and tail (sample < 50) classes and evaluate the accuracy of the different classes separately. 

%To evaluate \tool, we use our long-tailed distribution dataset extracted from Fan \et in vulnerability type classification and three widely-studied benchmark datasets in vulnerability detection. The results demonstrate that the \tool outperforms all the baseline methods in vulnerability prediction. Specifically, \tool achieves an accuracy of 64.01\% in vulnerability type classification, which is 25.75\% and 23.61\% absolute improvement over the best-performing approach. In the vulnerability detection task, \tool also improves the F1 score performance of 8.36\%, 23.74\%, and 63.7\% over the best-performing approach on three datasets.

%\item We systematically study existing approaches in vulnerability type classification and identify several problems with the long-tailed data distribution.
The major contributions of this paper are as follows:
\begin{enumerate}
\item We are the first to approach to explore
% overcome 
the long-tailed classification of vulnerable functions.
% distribution issue of the software vulnerability classification task.
\item We propose \tool, a long-tailed software
vulnerability type classification approach, including: 1) a vulnerability representation learning module for
% overcoming 
alleviating the over-smoothing issue of GNNs and enhancing the vulnerability representations; and 2) an adaptive re-weighting module that involves a novel training objective for balancing the weights of different types.
% to learn the tail class representation for alleviating the data imbalance.
\item Extensive experiments show the effectiveness of \tool in vulnerability type classification and the efficacy of vulnerability representation learning module in vulnerability detection.
% We perform a large-scale evaluation of \tool on vulnerability type classification, and the results demonstrate the effectiveness of \tool. We further  evaluate the efficacy of the vulnerability representation learning module in \tool on vulnerability detection. %\gsz{[we are the first to investigate fined-gained vulnerability classification problem?]}
%From the perspective of data's long-tailed distribution, we propose a novel loss function, Bilateral-branch Temperature Loss.
%\item Our study reveals several crucial problems and challenges in vulnerability type classification models based on the data which shows the long-tailed distribution, and conducts a comparison with previous methods leveraged to overcome the long-tailed problem. We verify the shortcomings of previous methods.
%\item ...
\end{enumerate}

The rest of this paper is organized as follows. Section~\ref{sec:background} describes the background. Section~\ref{sec:architecture} details the two components in the proposed framework of \tool, including the vulnerability representation learning module and adaptive re-weighting module. Section~\ref{sec:evaluation} describes the evaluation methods, including the datasets, baselines, implementation and metrics. Section~\ref{sec:experimental_result} presents the experimental results. Section~\ref{sec:discussion} discusses some cases and threats to validity. Section~\ref{sec:conclusion} concludes the paper.

\section{Background}
\label{sec:background}
% \yc{reconsider this section}

\subsection{Graph Neutral Networks}% and Over-smoothing}
\label{sec:oversmooth}
Graph Neural Networks (GNNs) have demonstrated superior performance at capturing the structural information of source code in the software vulnerability detection task~\cite{DBLP:conf/nips/ZhouLSD019/,DBLP:journals/tse/ChakrabortyKDR22/, DBLP:conf/sigsoft/Li0N21/}. The widely-used GNN methods include Graph Convolutional Network (GCN)~\cite{kipf2016gcn}, Graph Attention Network (GAT)~\cite{DBLP:journals/corr/abs-1710-10903/GAT}, GGNN. The two-layer of GCN model can calculate as:
% mining graph data structures for software engineering tasks, such as code classification~\cite{DBLP:journals/corr/abs-1903-03804/cc1, combefis2022automated/cc2}, code clone detection~\cite{DBLP:journals/corr/abs-2002-08653/ccd1,DBLP:journals/ijseke/JiLZ21/ccd2,DBLP:journals/jss/LeiLLAK22/ccd3}, \etc 
% One widely used method is Graph Convolutional Network (GCN)~\cite{kipf2016gcn}, a two-layer GCN model can calculate as:
\begin{equation}
\label{gnn}
H = softmax(\hat{\tilde{A}}ReLU(\hat{\tilde{A}}XW_{0})W_{1})
\end{equation}
where $\hat{\tilde{A}}$ %\gsz{[ $\hat{\tilde{A}}$ in equation, make it consistent]} 
denotes the adjacency matrix and $W_{0}, W_{1}$ are trainable weight matrices. $X$ and $H$ denote the initial and output node representations respectively. 
With the two layers, only neighbors in the two-hop neighborhood are considered for the node representation learning. However, due to the deep levels in the code structure graph~\cite{graphcodebert, DBLP:conf/icse/ZhangWZ0WL19/ASTNN}, a two-layer GNN is hard to well capture the code node representations. Thus, the GNNs tend to learn similar node representations during the propagation steps~\cite{DBLP:conf/icml/XuLTSKJ18/XUet}, i.e., the \textit{over-smoothing} issue~\cite{bottleneck}, failing to capture the vulnerability patterns for different types. To mitigate the over-smoothing issue and distinguish the node representations, a vulnerability representation learning module is proposed by improving the node feature propagation method.

\subsection{Long-tailed Learning Methods}
%Previous vulnerability detection methods, such as Reveal~\cite{DBLP:journals/tse/ChakrabortyKDR22/} and IVDetect~\cite{DBLP:conf/sigsoft/Li0N21/}, evaluate whether a function contains a vulnerability. These methods face a large class imbalance during training, as the number of samples of functions with vulnerabilities is much smaller than the number of samples that do not contain vulnerabilities. 
% The imbalanced sample distribution is even worse in the problem of vulnerability type classification. Overall, data for different vulnerability types follow %unexpected 
% long-tailed distributions~\cite{DBLP:journals/corr/abs-2110-04596/longtail}, where the numbers of instances for different classes are seriously imbalanced. 

In real-world scenarios, data
% samples 
typically exhibit a long-tailed distribution~\cite{DBLP:journals/corr/abs-2110-04596/longtail}, where a small portion of classes contain massive samples while the others
% has a massive sample  but the other classes 
are associated with only a few samples. 
% With long-tailed data distributions, the existing way of training models leads to two problems: \wxc{(1) it is difficult to effectively learn the representation of each vulnerability type due to the serious imbalance. A large number of head samples can make the model overly trust the samples in the head classes, leading to model degradation and difficulty in learning discriminative vulnerability representations overall. (2)
%  lack of tail-class samples makes it challenging to train models for tail-class classification because the number of samples for most vulnerability types (tail classes) is too small for the model to learn useful learning \gsz{features of} %signals from
%  these classes; }
Long-tailed learning methods have been
% are 
widely studied in the computer vision field~\cite{DBLP:conf/cvpr/SzegedyVISW16/, DBLP:journals/corr/abs-2110-04596/longtail, DBLP:conf/cvpr/TanWLLOYY20/reweight1}.
% The long-tailed distribution learning method has been widely studied in computer vision~\cite{DBLP:conf/cvpr/SzegedyVISW16/, DBLP:journals/corr/abs-2110-04596/longtail, DBLP:conf/cvpr/TanWLLOYY20/reweight1}.
There are some popular methods to alleviate long-tailed problems, such as class-balanced strategies~\cite{DBLP:conf/cvpr/CuiJLSB19/, DBLP:conf/iccv/LinGGHD17/} and smoothing strategies~\cite{DBLP:conf/cvpr/SzegedyVISW16/, DBLP:conf/cvpr/ZhongC0J21/}. For example, 
% a widely used method is 
one popular method Focal Loss~\cite{DBLP:conf/iccv/LinGGHD17/}
% , which 
adjusts the standard cross-entropy loss~\cite{DBLP:conf/nips/ZhangS18/celoss} by reducing
% to reduce 
the learning weight for well-classified samples and focuses more on misclassified samples during model training, calculated as:
% Specifically, it calculates as:
\begin{equation}
\label{fl}
{\mathcal{L}}_{FL} = - \sum_{i=1}^{n}(1-\hat{y_{i}})^{\gamma}y_{i}log(\hat{y_{i}})
\end{equation}
where $y_{i}$ and $\hat{y_{i}}$ denote the label distribution in the ground truth and the predicted output, respectively.
% predicted by the classifier. 
$\gamma$ is the modulating factor for focusing on misclassified samples.

Label smooth cross-entropy loss~\cite{DBLP:conf/cvpr/SzegedyVISW16/} is also a widely used long-tailed solution. 
% strategy. 
It uses smoothing strategies to encourage the model to be less confident in head classes:
\begin{equation}
\label{LSCE}
{\mathcal{L}}_{LSCE} = - \sum_{i=1}^{n}log(\hat{y_{i}})\left((1-\epsilon)y_{i} + \epsilon \delta_{i}\right )
\end{equation}
where $\epsilon$ denotes a smoothing
% weight 
parameter, and $\delta_{i}$ denotes the uniform distribution to smooth
% change 
the ground-truth distribution $y_{i}$.

Although the
% all 
existing methods alleviate the problem of long-tailed distributions in computer vision, no studies have explored the performance of these methods in software vulnerability classification. To fill the gap, we experimentally evaluate the popular long-tailed learning methods, and propose a novel adaptive learning method for vulnerability type classification.
%Although all existing approaches propose a model to report good performance, 
% none of them incorporate the characteristics of vulnerability type classification, e.g., large differences in the difficulty of classification between classes, difficulty in learning vulnerability representations, etc.
% Although all existing methods propose a model to report good performance, it does not focus on the problem of class imbalance in vulnerability type classification under long-tailed distributions. 
 % In this paper, we propose an adaptive learning module, which shifts the models’ learning weights in different classes and training phases for alleviating the data imbalance.

\section{Proposed Framework}
\label{sec:architecture}
\begin{figure*}[t]
	\centering
    \includegraphics[width=1.0\textwidth]{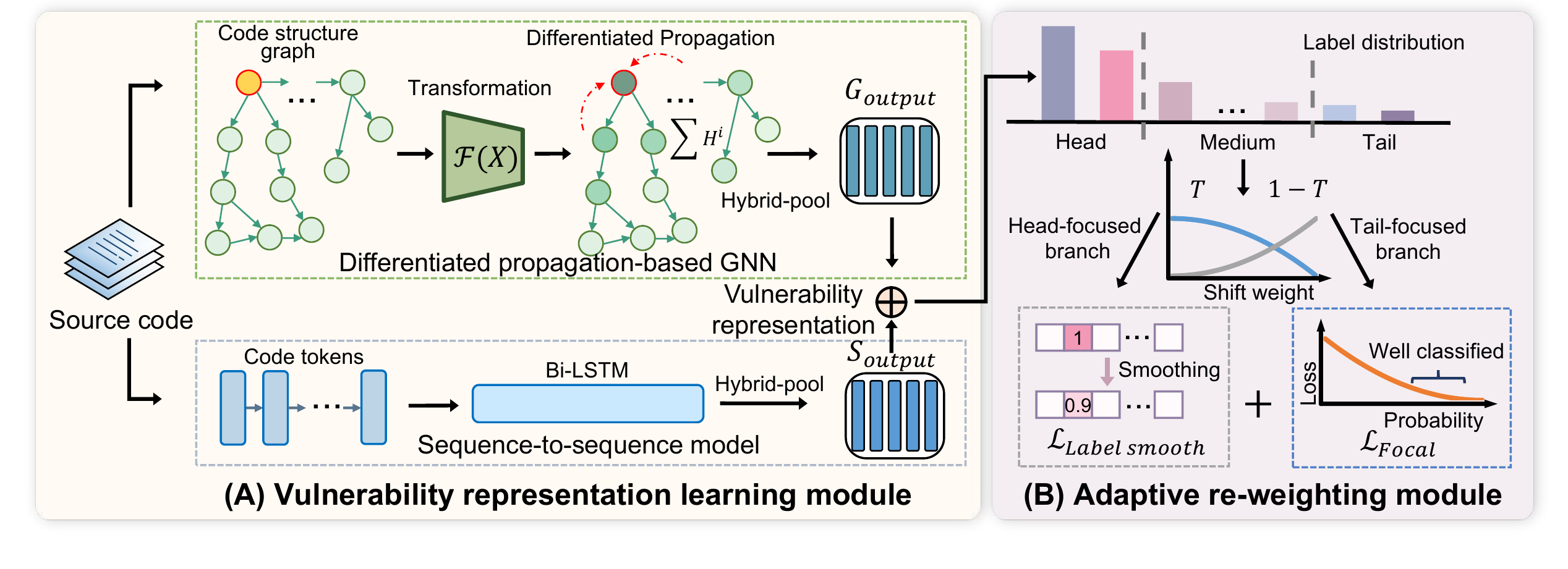}
	\caption{The architecture of \tool, which mainly contains two components: (A) a vulnerability representation learning module, and (B) an adaptive re-weighting module. }%Hybrid-pool layer involves a maximum pooling operation~\cite{DBLP:journals/nca/YueXY19/maxpool} and an average pooling operation~\cite{DBLP:conf/icml/BoureauPL10/avepool} layer.}
	\label{architecture}
\end{figure*}
In this section, we first formulate
% review 
the long-tailed problem
% formulation 
and then describe the overall framework of the proposed \tool.
As shown in Figure~\ref{architecture}, \tool consists of two main
components: (1) a vulnerability representation learning module for enhancing the vulnerability representations; and (2) an adaptive re-weighting module
that involves a novel training objective for balancing the weights of different types. %\yun{[the sentence needs to be rephrased???]}.
\subsection{Problem Formulation}
% We first 
The problem of
% define 
the long-tailed vulnerability type classification is formulated below.
% settings. An 
We denote the unbalanced dataset with training instances
% is 
collected from the real world as $D =\left \{x_{i}, y_{i} \right\}$, where $x_{i}$ denotes a vulnerable function in raw source code and $y_{i}$ denotes its corresponding vulnerability type. Assume that all classes are ordered by cardinality, i.e., if $N_{a} \ge N_{b}$ when the class index $a < b$, where $N_{a}$ indicates
% is 
the number of training samples for the class $a$. According to the typical
% previous 
division ratio in long-tailed scenario~\cite{DBLP:conf/cvpr/ZhongC0J21/, DBLP:conf/cvpr/0002021}, we classify the vulnerable types into three classes: head (>200 samples per type), medium (50-200 samples per type), and tail (<50 samples per type).
%, which account for 85\%, 10\% and 5\%, respectively. 
As a result, the head, medium, and tail classes contain 12, 11, and 46 types, respectively.
%The head (>1000 samples per class), medium (200-1000 samples per class), and tail (<200 samples per class) classes contain
% \wxc{we divide the classes into the head, medium, and tail classes respectively accounting for 85\%, 10\% and 5\%,}
%we divide the classes into three groups, 
%with the number of class samples greater than 200 denoted as head classes, the number of samples less than 50 denoted as tail classes, and the remaining classes referred to as medium classes.
% In addition, we group the number of samples with $N_{a}<20$ into a class because it is difficult to detect cases with too few test samples. 
The goal of \tool is to learn a mapping, $f: x_{i} \mapsto y_{i}$, $y_{i}$ denotes the label distribution, to predict the class of vulnerabilities function. Typically, we use the Cross-Entropy (CE) loss~\cite{DBLP:conf/nips/ZhangS18/celoss} function as follows:
\begin{equation}
\label{ce}
{\mathcal{L}}_{CE} = - \sum_{i=1}^{n}y_{i}log(\hat{y_{i}})
\end{equation}
where $n$ denotes the number of categories and $\hat{y_{i}}$ denotes the output predicted by the model. %This can lead to a model being over-biased on the head classes and neglecting to focus on the tail classes. In this paper, we propose \tool to improve the focus on the tail classes.
\subsection{Vulnerability Representation Learning Module}
In this section, we elaborate on the proposed
% part, we propose a 
vulnerability representation learning module, which involves differentiated propagation-based GNN for capturing the structural information of source code,
% in the source code 
and combines the sequence-to-sequence approach
% a sequence branch 
for further enhancing the vulnerability representations.
\subsubsection{Differentiated propagation-based GNN}

%In \tool, we focus on structural information in programming languages. 
%Programming languages differ from natural languages in that they have inherent syntactic and semantic features such as syntactic structure (AST), data and control dependency information (DFG and CFG), etc.  Instead of compressing information into a flat vector, a graph neural network (GNN) learns graph structure data based on an information diffusion mechanism that updates node states based on the connectivity of the graph to retain important information, i.e. topological dependency information. Therefore, 
Following the prior studies~\cite{DBLP:conf/nips/ZhouLSD019/}, we extract the same code structure graph of source code for vulnerability representation. To alleviate the over-smoothing issue of GNN, we design a differentiated propagation method to distinguish node representations.

% In vulnerability detection, previous approaches have typically used the GNN model and demonstrate that GNNs can effectively capture vulnerability patterns. They extract the code structure graph from the source code and directly adopt the GNN model, which updates the state of nodes based on the connectivity of the graph to preserve important information, i.e. topological dependencies in the graph. 
%However, directly adopting previous vulnerability detection approaches tends to result in poor performance.
%In each GNN layer, each node vector is updated with two processes: feature transformation (\ie MLP layer) and propagation.
% As discussed in Section \ref{sec:oversmooth}, we know that the GNN converges to the limit distribution of the random walk as the number of layers increases. The limit distribution is a property of the whole graph, but the vulnerability node is only contained in a part of the graph that does not take into account the starting (root) node of the random walk. As a result, the performance of GNN for vulnerability type classification at higher layer counts (or multiple propagation/linear transformations) is degraded. \wxc{In this paper, we propose a differentiated propagation GNN (\ie graph branch) for learning the vulnerability representation, which involves a differentiated propagation to distinguish node representations.}

% Following the previous work~\cite{DBLP:conf/nips/ZhouLSD019/}, we use 
We denote the code structure graph $\mathcal{G}(\mathcal{V},\mathcal{E})$ as the input, where
$\mathcal{V}$ denotes the set of nodes $v$ and $\mathcal{E}$ denotes the set of edges $e$ in the graph.
%In contrast to all previous GNNs on code vulnerability detection, 
The differentiated propagation-based GNN is decoupled into two steps, including node feature transformation and propagation. The representation for each node $v$ is first initialized as a 128-dimensional vector by the Word2Vec~\cite{DBLP:journals/nle/Church17/word2vec} model. To enhance the representation of each node vector, the feature transformation step then encodes the vector feature based on a Gated Recurrent Unit (GRU) layer:
% We decouple the differentiated propagation GNN into feature transformation and propagation steps.  
%The feature transformation stage is designed to better capture the semantic information of the source code. 
% Specifically, the transformation feature $H^{(0)}$ is defined for each node as follows:
\begin{equation}
\label{initial}
H^{(0)} = {\mathcal{F}}\left (X \right)
\end{equation}
where $X$ is the initial node representation of node $v$, and $\mathcal{F}$ denotes a GRU layer.
% Each node $v$ is initialized by the Word2Vec~\cite{DBLP:journals/nle/Church17/word2vec} model and encoded into a 128-dimensional vector. $\mathcal{F}$ denotes a GRU layer and operates on each node’s vector independently. %In addition, the subsequent steps does not have a non-linear transformation (\ie a MLP layer), which improves the efficiency of the calculation.

In the node feature propagation step, considering that the node representations in GNNs tend to converge to a similar during the propagation,
% converges to the limit distribution of the random walk
we propose
% a differentiated propagation method 
to involve the initial node representations. The node representation $H^{l}$ at the $l$-th layer is defined as:
% node vector in the graph, \ie not getting lost in the limit distribution of the random walk. 
% The differentiated propagation is calculated as below: 
\begin{equation}
\label{propagation}
H^{l} = \frac{1}{l} \sum_{i=0}^{l-1} \left ( \left ( 1-\alpha \right ) \left ( {\widetilde{D}^{-\frac{1}{2}}}{\widetilde{A}}{\widetilde{D}^{-\frac{1}{2}}}\right )^{i} H^{0} + \alpha H^{(0)}  \right )
\end{equation}
% $l$ is the layer for model aggregation and 
where $H^{(0)}$ is the transformed feature representation computed by Equation~(\ref{initial}).
% node feature after transformation steps.
$\widetilde{A}$ is the adjacency matrix of the code structure graph with self-connections and $\widetilde{D}$ is the degree matrix of $\widetilde{A}$. $\alpha$ denotes a teleport hyperparameter, which determines the importance of the initial node features
% vector 
during the propagation. Finally, the graph representation $G_{output}$ of the code structure graph $\mathcal{G}$
% graph branch output 
is calculated through a hybrid pooling layer:
% the Hybrid-pool layer as follows:
\begin{equation}
G_{output} = \mathcal{C}\left(AvgPool(H_{v}) + MaxPool(H_{v})\right ), \forall{v} \in \mathcal{V}
\end{equation}
% Compared to using a single operation, 
where $AvgPool(\cdot)$ and $MaxPool(\cdot)$ 
indicate the average pooling operation~\cite{DBLP:conf/icml/BoureauPL10/avepool} and maximum pooling operation~\cite{DBLP:journals/nca/YueXY19/maxpool}, for better capturing the local information and global information of the code structure graph ~\cite{DBLP:conf/eccv/WooPLK18/cbam}, respectively. $\mathcal{C}(\cdot)$ denotes two Multi-Layer Perceptron (MLP) layers.
% MaxPool~\cite{DBLP:journals/nca/YueXY19/maxpool} can better capture the key nodes of the source code, while AvePool~\cite{DBLP:conf/icml/BoureauPL10/avepool} can better capture the global information of the source code. Here we use both MaxPool and AvePool to enhance the model learning capacity. $\mathcal{C}$ is two MLP layers to perform classification.
\subsubsection{Sequence-to-sequence model}
To further enhance the vulnerability representations,
% in code snippets, 
we involve a sequence-to-sequence model to capture the semantics information.
% (called sequence branch) 
% in the vulnerability representation learning module. 
% Compared to graph branches, the sequence model is more likely to detect different types of vulnerabilities in some cases, due to the fact that the same type of vulnerability may have different structure information (\ie vulnerability trigger paths). 
Specifically, given a code snippet $S$ and its composed code token sequence
% source code tokens $t_{i}$ in 
$S = \{t_{1},...,t_{i},...,t_{n} \}$, where $t_{i}$ denotes the $i$-th token and $n$ denotes the total number of tokens, the sequence representation
% sequence branch output 
$S_{output}$ is calculated as follows:
\begin{equation}
S_{output} = \mathcal{C}\left(AvgPool(s_{t_{i}}) + MaxPool(s_{t_{i}})\right ), \forall{t_{i} \in S}
\end{equation}
\begin{equation}
s_{t_{1}},...,,s_{t_{n}} = \left(\overrightarrow{LSTM}\left(t_{1},...,t_{n} \right) \right)||\left({ \overleftarrow{LSTM}\left(t_{1},...,t_{n}\right)}\right)
\end{equation}
where
% $n$ denotes the number of tokens in a code snippet and 
the symbol $||$ is the concatenation operation. $\overrightarrow{}$ and $\overleftarrow{}$ denote the Bi-directional Long Short-Term Memory (LSTM)~\cite{DBLP:BILSTM/} operations respectively.

Finally, the representation for each vulnerable code is computed by combining the two outputs:
\begin{equation}
    {\mathcal{O}}= G_{output} + S_{output}.
\end{equation}
% Finally, we add the vectors produced from the graph branch and sequence branch ${\mathcal{O}}= G_{output} + S_{output}$ for each code snippet.
% In the vulnerability detection task, we use a classifier and CE loss to report whether a code snippet is vulnerable or not.
% \begin{equation}
% \label{initial}

% \end{equation}

\subsection{Adaptive Re-weighting Module}
\label{sec:adaptive}
% In previous work, it has always been to concentrate on mapping source code to CWE-IDs under balanced conditions. However, the data distribution of different vulnerability types conforms to the long-tailed data distribution. As a result, current approaches have always focused excessively on the head classes with a high number of vulnerabilities. These methods can be easily biased towards head classes with massive training data, leading to neglecting the tail classes with a small number. 
The adaptive re-weighting module is proposed to better learn the representations for vulnerability types with different numbers of samples. A novel training objective is designed to adjust the
learning weights for different types of vulnerabilities according to the training epochs
and a number of associated samples.
% In this paper, we propose an adaptive re-weighting module, \wxc{which takes care of both head-classes learning and tail-classes learning simultaneously. We design a novel training loss to adjust the learning weights for different types of vulnerabilities according to the training epochs and number of associated samples.} 

% Here

Before the module design, we first investigate the popular long-tailed learning methods in vulnerability type classification, such as class-balanced strategies~\cite{DBLP:conf/cvpr/CuiJLSB19/, DBLP:conf/iccv/LinGGHD17/} and smoothing strategies~\cite{DBLP:conf/cvpr/SzegedyVISW16/, DBLP:conf/cvpr/ZhongC0J21/}.
We experimentally evaluate the performance of these methods.  
And the results are illustrated in Table~\ref{e2_multi} and will be detailed in Section~\ref{E_multi}.
We find that focal loss~\cite{DBLP:conf/iccv/LinGGHD17/} adds a modulating factor to focus more on tail samples. The label smooth CE loss~\cite{DBLP:conf/cvpr/SzegedyVISW16/} uses the smoothing strategies to reduce the focus on head classes.
The experiment results also demonstrate that focal loss performs better for classifying tails,
% tail classification, 
while the label smooth CE loss helps to improve the performance of head classes. %\yun{[briefly explain the reason??]}
% Specifically, 
Based on the observation, we propose a novel training objective that involves two training branches, i.e., a
% to involve two branches, which include the 
``\textit{tail-focused branch}'' for tail classification and ``\textit{head-focused branch}'' for head classification.

Specifically, these two branches
% use the same network structure model and 
take $\mathcal{O} = \{\hat{y_{i}}|i = 1,2,..,n\}$ as input, where $n$ denotes the number of types
% categories 
and $\hat{y_{i}}$ denotes the output
predicted by the model. The adaptive re-weighting method
% module 
$\mathcal{L}$ is calculated as follows:
\begin{equation}
\mathcal{L} = T \cdot \mathcal{L}_{FL} \left ({\mathcal{O}} \right) + (1-T) \cdot \mathcal{L}_{LSCE} \left ({\mathcal{O}} \right)
\end{equation}
where $\mathcal{L}_{FL}$ denotes the focal loss which focuses more on learning
% is used to learn 
% the 
the representations of tail classes.
% Because the tails have a smaller sample size and are more difficult to classify. 
$\mathcal{L}_{LSCE}$ denotes the label smooth CE loss focuses more on learning the representations of head classes. $T$ denotes a learning weight for
% between 
the two branches, which are calculated according to the training epochs:

% , which focuses on the head class. It is the annotation of vulnerabilities that may have some label noise~\cite{DBLP:journals/corr/abs-2301-05456/noise}, and reducing the recognition of head classes can improve model performance.
% First, w
%\yun{Specifically,} we design \yun{an adaptively learning strategy for shifting the 

% a learning strategy for shifting the shifting weight $T$ between two branches, which the shifting weight $T$ is calculated as:
\begin{equation}
T = 1 - \left ( \frac{E_{now}}{E_{max}}\right)^{2}
\end{equation}
where $E_{now}$ and $E_{max}$ denote the current training epoch and the total number of training epochs, respectively. The design of $T$ is based on the assumption that the representations of head classes are more easily learnt in the long-tailed scenario~\cite{DBLP:conf/cvpr/ZhouCWC20/bbn}. In the early training stage (i.e, with a smaller $E_{now}$), the design will enable the model to focus on learning the representations of tail classes. As the training epoch increases, $T$ will gradually decrease, shifting the model' learning focus to head classes. $T$ ensures that both branches are updated continuously throughout the training process, 
% which can 
avoiding learning conflict between the two branches.

\section{EXPERIMENTAL Setup}
\label{sec:evaluation}
\begin{table}[]
\centering
\setlength{\tabcolsep}{1.5mm}
\renewcommand{\arraystretch}{1.1}
\caption{The specific vulnerability types and their corresponding proportion and group of the types of vulnerabilities in this paper. Classes with a sample size of less than 20 are grouped together in the Remain class. ``None type'' means the vulnerability is not classified into any class. As these vulnerabilities exist in the real world as well, they are also considered to be a vulnerability type.}
\begin{tabular}{cr|c|cr|c}

\toprule
Types & Ratio & Group                   & Types & Ratio & Group                   \\ \midrule
CWE-119          & 19.94\%   & \multirow{12}{*}{Head}  & CWE-415          & 0.76\%    & \multirow{7}{*}{Medium} \\  
None type        & 19.85\%   &                         & CWE-732          & 0.62\%    &                         \\  
CWE-20           & 10.71\%   &                         & CWE-404          & 0.58\%    &                         \\  
CWE-399          & 6.90\%    &                         & CWE-79           & 0.52\%    &                         \\  
CWE-125          & 5.86\%    &                         & CWE-19           & 0.52\%    &                         \\  
CWE-264          & 4.76\%    &                         & CWE-59           & 0.49\%    &                         \\  
CWE-200          & 4.72\%    &                         & CWE-17           & 0.48\%    &                         \\  \cline{4-6} 
CWE-189          & 3.16\%    &                         & CWE-400          & 0.45\%    & \multirow{8}{*}{Tail}   \\  
CWE-416          & 3.09\%    &                         & CWE-772          & 0.43\%    &                         \\  
CWE-190          & 2.88\%    &                         & CWE-269          & 0.36\%    &                         \\  
CWE-362          & 2.61\%    &                         & CWE-22           & 0.33\%    &                         \\  
CWE-476          & 2.02\%    &                         & CWE-369          & 0.32\%    &                         \\ \cline{1-3}
CWE-787          & 1.86\%    & \multirow{4}{*}{Medium} & CWE-18           & 0.32\%    &                         \\  
CWE-284          & 1.66\%    &                         & CWE-835          & 0.32\%    &                         \\  
CWE-254          & 1.15\%    &                         & Remain class     & 1.57\%    &                         \\   
CWE-310          & 0.88\%    &                         &     &                         \\ \bottomrule
\end{tabular}
\label{fan_sta}
\end{table}

\subsection{Research Questions}
In order to evaluate \tool, we answer the following research questions:
\begin{enumerate}[label=\bfseries RQ\arabic*:,leftmargin=.5in]
    \item How well does \tool perform in classifying vulnerability types under the long-tailed distribution?
    \item How effective is the vulnerability
representation learning module in vulnerability detection?
    
    \item What is the impact of different components on the type classification performance of \tool?
    \item What is the influence of hyper-parameters on the performance of \tool?
    % hyperparameter tuning influence of \tool when detecting source code vulnerability?}
\end{enumerate}

\subsection{Datasets}

\textbf{Vulnerability type classification.} To obtain the labels of vulnerabilities for type classification, we extract a new dataset from Fan \et~\cite{DBLP:conf/msr/FanL0N20/}, which consists of different types of vulnerabilities from 2002 to 2019 and provides the \emph{Common Weakness Enumeration IDentifier} (CWE ID)~\cite{CWEID} for each vulnerable function. 
% We obtain 10667 vulnerable functions with 91 different kinds of CWE-ID. Categories with a sample size of less than 20 are grouped together.
%Our study extracts a new dataset from the Fan \et~\cite{DBLP:conf/msr/FanL0N20/} for the reason that the dataset consists of 91 different types of vulnerabilities from 2002 to 2019, each of which is uniquely identified by a \emph{Common Weakness Enumeration IDentifier} (CWE ID)~\cite{CWEID}. To satisfy the vulnerability classification setting, we obtain 10667 vulnerable functions with 69 different kinds of CWE-ID. Samples with a sample size of less than 10 are grouped together. 
We obtain a total of 10667 vulnerable functions with 91 different kinds of CWE-ID. Specifically, as shown in Figure~\ref{fan_sta}, we demonstrate the group and the proportion of head, medium, and tail classes in our dataset. The percentage of samples in the head, medium, and tail classes amount to 86.50\%, 9.52\%, and 4.00\% respectively. Moreover, we group categories with a sample size of less than 20 as a new class named \emph{Remain class}.

%The FFMPeg+Qemu and Reveal datasets do not contain type information of vulnerability.
%\lf{In Figure~\ref{fan_all}, we demonstrate the proportion of CWE IDs in the dataset. The majority of the pie chart segments are occupied by the head classes. In contrast, the tail classes segments in the pie chart mostly only occupy around 1\% of the total proportion. For instance, the proportion of CWE-119 (belonging to the head class) is 20\%, which is 20 times greater than that of CWE-19 (belonging to the tail class)}
% Given the demonstration of the long-tailed distribution displayed in Figure~\ref{cweid}, we have conducted a further categorization of CWE IDs into three classifications: head, medium, and tail. CWE IDs with a number of samples greater than 200 will be classified as head class. Those between 50 and 200 will be classified as medium class. CWE IDs with a number of samples less than 50 will be classified as tail class.
%In \tool, we need to transfer the source code into a graph. However, in light of certain limitations such as graph size, some samples are removed.

\textbf{Vulnerability detection.}
Our study employs three representative vulnerability datasets:
FFMPeg+Qemu~\cite{DBLP:conf/nips/ZhouLSD019/}, Reveal~\cite{DBLP:journals/tse/ChakrabortyKDR22/}, and Fan \et~\cite{DBLP:conf/msr/FanL0N20/}. The FFMPeg+Qemu dataset, previously utilized in Devign{~\cite{DBLP:conf/nips/ZhouLSD019/}} with +22K data, exhibits a vulnerability rate of 45.0\% among all instances. The Reveal dataset consists of +18K instances, and the proportion of vulnerable to non-vulnerable data is 1:9.9. The Fan \et dataset includes a total of 188,636 C/C++ function samples, among which 5.7\% are vulnerable. We do not use FFMPeg+Qemu~\cite{DBLP:conf/nips/ZhouLSD019/} and Reveal~\cite{DBLP:journals/tse/ChakrabortyKDR22/} datasets for vulnerability type classification since these two datasets do not contain type information of vulnerability.

\subsection{Baseline Methods}
In vulnerability detection task, we compare \tool with three sequence-based methods~\cite{DBLP:conf/ndss/LiZXO0WDZ18/,DBLP:conf/icmla/RussellKHLHOEM18/,DBLP:journals/tdsc/0027ZX0ZC22/} and three state-of-the-art graph-based methods~\cite{DBLP:conf/nips/ZhouLSD019/,DBLP:journals/tse/ChakrabortyKDR22/,DBLP:conf/sigsoft/Li0N21/}.
\begin{enumerate}
\item \textbf{VulDeePecker}~\cite{DBLP:conf/ndss/LiZXO0WDZ18/}: VulDeePecker embeds the data flow dependency to construct the code slices. Then, it adopts the BiLSTM to detect buffer error vulnerabilities and resource management error vulnerabilities. %A sequence-based method that combines data flow analysis and BLSTM to detect buffer error vulnerabilities and resource management error vulnerabilities.
\item \textbf{Russell \et}~\cite{DBLP:conf/icmla/RussellKHLHOEM18/}: Russell \et involves the Convolutional Neural Network (CNN), integrated learning and random forest to vulnerability detection.
\item \textbf{SySeVR}~\cite{DBLP:journals/tdsc/0027ZX0ZC22/}: SySeVR constructs the program slices by combining control flow dependency, data flow dependency, and code statement. It embeds program slices as input and uses Recursive Neural Networks (RNN).%The approach which combines both control and data flow in the code representations using Recurrent Neural Networks (RNNs).
\item \textbf{Devign}~\cite{DBLP:conf/nips/ZhouLSD019/}: Devign constructs code structure graph from functions and leverages Gated Graph Neural Network (GGNN) for classification.%A typical graph-based method that uses GGNN and builds graphs combining Abstract Syntax Tree (AST), Control Flow Graph (CFG), Data Flow Graph (DDG) and Natural Code Sequence (NCS).
\item \textbf{Reveal}~\cite{DBLP:journals/tse/ChakrabortyKDR22/}: Reveal leverages code property graph (CPG) as input and also adopts GGNN for extracting features. Then it utilizes a multi-layer perceptron for vulnerability detection.%The graph-based method that uses a combination of GGNN, Multi-Layer Perceptron (MLP) and triplet Loss to construct the model, where the code property graph is used as input representation. 
\item \textbf{IVDetect}~\cite{DBLP:conf/sigsoft/Li0N21/}: IVDetect constructs Program Dependency Graph (PDG) and designs the feature attention GCN for vulnerability detection.%This method leverages a model which is based on the graph convolutional network (GCN), and the input features of this model take contexts via data and control dependencies into account.
\end{enumerate}

In vulnerability type classification,
we compare \tool with graph-based methods: Devign and Reveal. These methods are designed for the software vulnerability detection task and achieve the best-performing results. We also use the focal loss~\cite{DBLP:conf/iccv/LinGGHD17/} and label smooth CE (LSCE) loss~\cite{DBLP:conf/cvpr/SzegedyVISW16/} to replace CE loss, which has been introduced in Section~\ref{sec:adaptive}.
Furthermore, we compare with label aware smooth loss~\cite{DBLP:conf/cvpr/ZhongC0J21/}, class-balanced loss~\cite{DBLP:conf/cvpr/CuiJLSB19/}, and class-balanced focal loss~\cite{DBLP:conf/cvpr/CuiJLSB19/} that mitigates the imbalance of long-tailed distribution.
% FL~\cite{DBLP:conf/iccv/LinGGHD17/} decreases the weight of well-classified samples in order to make the model focus on poorly-classified samples during training.
% LSCE~\cite{DBLP:conf/cvpr/SzegedyVISW16/} smoothe s the original ground-truth distribution to relieve the over-confidence problem of the model in the head class.
Label aware smooth loss adds a probability distribution factor based on LSCE, which considers the predicted probability distributions of different classes. Class-balanced loss and class-balanced focal loss are modified from CE loss and focal loss, respectively. They add a weight inversely related to the number of class samples to tackle long-tailed distribution.
%Label aware smooth CE~\cite{DBLP:conf/cvpr/ZhongC0J21/} combines takes the fact that different classes have different predicted probability distributions into account and combines with a label smoothing method.、

%Class-balanced loss~\cite{DBLP:conf/cvpr/CuiJLSB19/} uses a weight that is inversely proportional to the number of classes using the effective numbers on the basis of CE loss. By incorporating this into the Focal Loss, we get the class-balanced focal loss~\cite{DBLP:conf/cvpr/CuiJLSB19/}.

\subsection{Implementation Details}
In the experiment section, we randomly split the dataset into training, validation, and testing sets by the number of classes in a ratio of 8:1:1. To ensure the fairness of the experiments, we use the same data split in all experiments. We train the vulnerability detection model for 100 epochs and the vulnerability type classification model for 50 epochs on a server with an NVIDIA GeForce RTX 3090.

We leverage Word2Vec~\cite{wv} to initialize the node representation in the graph branch and the token representation in the sequence branch, where both vectors have a dimension of $d = 128$. The number of layers in the graph branch is $L =16$ and the limitation and the source code length in the sequence branch is $n = 512$. The dimension of the hidden vector in the sequence branch is set as 512.
%The dimension of the initial graph representation and sequence representation is set as Word2Vec and has a dimension of 128.  The number of layers in the graph branch is 16. The source code length of the sequence branch is limited to 512 tokens and the hidden layer dimension is set as 512.

\subsection{Performance Metrics}

We use the following four widely-used performance metrics in our evaluation:

% Let \emph{TP} denote the number of truly vulnerable samples which are detected as vulnerable, \emph{FP} denote the number of truly non-vulnerable samples which are detected as vulnerable, \emph{TN} denote the number of truly non-vulnerable samples which are detected as non-vulnerable, and \emph{FN} denote the number of truly vulnerable samples which are detected as non-vulnerable.

% \textbf{Precision:} 
% The metric $Precision = \frac{TP}{TP+FP}$ is used to measure the proportion of correctly identified vulnerable samples among all the samples detected as vulnerable by the model.

% \textbf{Recall:} 
% The metric $Recall = \frac{TP}{TP+FN}$ measures the ratio of the samples that are detected by the model as vulnerable among entire the actual vulnerable samples.

% \textbf{F1 score:} 
% The metric $F1 score = \frac{2*Precision*Recall}{Precision+Recall}$, which is the harmonic mean of \emph{Precision} and \emph{Recall}. 

% \textbf{Accuracy:} 
% The metric $ALL\_A = \frac{TP+TN}{TP+FP+TN+FN}$ measures the proportion of the entire samples which are detected correctly by the model. The calculations of metrics $Head\_A$, $Mid\_A$, $Tail\_A$ are as the same, where $Head\_A$, $Mid\_A$ and $Tail\_A$ separately measure the proportion of the head, medium, and tail CWE ID samples which are detected correctly.

\textbf{Accuracy:} 
$Accuracy = \frac{TP+TN}{TP+FP+TN+FN}$. Accuracy is the proportion of correctly classified instances to all instances. \emph{TP} is the number of true positives, \emph{TN} is the number of true negatives. and $TP+FP+TN+FN$ represents the number of all instances. To evaluate the classification performance of different classes of vulnerabilities, we use $Head$, $Medium$, $Tail$ to denote the proportion of the head, medium, and tail CWE samples that are detected correctly, respectively. %\lf{For evaluating the comparison results between \tool and the baselines in vulnerability type classification conveniently, we further define three metrics for the different group's accuracy: $Head$, $Medium$, $Tail$ are used to separately measure the proportion of the head, medium, and tail CWE ID samples which are detected correctly. }

\textbf{Precision:} 
$Precision = \frac{TP}{TP+FP}$. Precision is the proportion of relevant instances among those retrieved. \emph{TP} is the number of true positives and \emph{FP} is the number of false positives.

\textbf{Recall:} 
$Recall = \frac{TP}{TP+FN}$. Recall is the proportion of relevant instances retrieved. \emph{TP} is the number of true positives and \emph{FN} is the number of false negatives.

\textbf{F1 score:} 
$F1 score = 2 *\frac{Precision*Recall}{Precision+Recall}$. F1 score is the geometric mean of precision and recall and indicates the balance between them.

\section{Experimental Results}
\label{sec:experimental_result}

 \subsection{RQ1: Evaluation on Vulnerability Types Classification}
 \label{E_multi}
 %RQ1: How well does \tool perform in classifying vulnerability types under the long-tailed distribution?

\begin{table*}[h]
\centering

\setlength{\tabcolsep}{1.1mm}
\renewcommand{\arraystretch}{1.1}

\caption{Under the long-tail data distribution, comparison results between \tool and the baselines in vulnerability type classification. VRL module denotes the vulnerability representation learning module.}
\resizebox{.97\textwidth}{!}{
\begin{tabular}{l|rrrr|rrrr|rrrr}
\toprule
\diagbox{Metrics(\%) }{Baseline} & \multicolumn{4}{c|}{Devign}        & \multicolumn{4}{c|}{Reveal}             & \multicolumn{4}{c}{VRL module in \tool}                \\
\midrule

Re-weighting                 & \multicolumn{1}{c}{Head}    & \multicolumn{1}{c}{Medium}  & Tail    & Accuracy  & Head    & Medium  & Tail    & Accuracy  & Head    & Medium  & Tail    & Accuracy       \\
\midrule

CE loss       & 38.26 & 7.35  & 23.08 & 34.99 & 39.39&42.65&46.15&39.83 & 60.61& 51.47& 38.46& 59.32\\
Label aware smooth loss & 37.14 & 25.00 & 23.08 & 35.70 & 38.26 & 42.65 & 46.15& 38.83 & \cellcolor{gray!45}63.50 & 52.94& 46.15& \cellcolor{gray!20}62.16 \\
Label smooth loss       & 37.78 & 4.41  & 23.08 & 34.28 & 39.07 & 42.65 & 46.15& 39.54 & \cellcolor{gray!45}63.50 & \cellcolor{gray!45}55.88 & 46.15& \cellcolor{gray!45}62.45 \\
Class-balanced loss    & 27.97 & 36.76 & 38.46 & 29.02 & 37.30 & 44.12 & 46.15& 38.12 & 60.45& \cellcolor{gray!20}54.41& \cellcolor{gray!45}53.85 & 59.74\\
Class-balanced focal loss  & 24.44 & 7.35  & 23.08 & 22.76 & 27.17&42.65&46.15&29.02       & 54.82& \cellcolor{gray!20}54.41& \cellcolor{gray!45}53.85 & 54.77\\
Focal loss           & 38.91&14.71&38.46&37.27 & 38.59 & 41.18 & 46.15& 38.98 & 61.09& 52.94&\cellcolor{gray!45}53.85&60.17\\
%Average               & 34.22 & 17.40 & 29.49 & 32.50 & 38.59 & 42.94 & 47.69& 39.17 & 61.28& 53.19& 47.44& 60.24\\

\midrule
%Our proposed         
Adaptive re-weighting module &39.27&	37.29&	50.00&	39.39 &   41.00 &47.06&46.15& 41.68&  \cellcolor{gray!70}\textbf{64.79}& \cellcolor{gray!70}\textbf{58.82}& \cellcolor{gray!70}\textbf{53.85}& \cellcolor{gray!70}\textbf{64.01}            \\  
\bottomrule
\end{tabular}}

\label{e2_multi}
\end{table*}

To answer this research question, we compare our approach with the previous methods of vulnerability type classification under long-tailed distribution.
\subsubsection{Compared with baselines}
As shown in Table~\ref{e2_multi}, the proposed \tool consistently outperforms all the baseline methods in the long-tailed distribution. Our approach achieves an accuracy of 64.01\%, which is an absolute improvement of 26.74\% and 24.18\% over Devign and Reveal, respectively. %25.75\% and 23.61\% absolute improvement over other baseline approaches. 
Compared with the baseline methods, we focus on modeling vulnerability patterns from two perspectives based on the graph branch and the sequence branch. The results show that our approach can learn vulnerabilities more efficiently from different perspectives. 
For the head, medium and tail classes, we can see that \tool achieves the best performance compared to the previous methods, with the improvement of 25.40\%, 16.17\% and 7.7\% respectively. It demonstrates that our approach is better at capturing the differences between the different vulnerability types. 
% We can observe that the \tool is overall more effective than baselines in head, medium and tail classes. 

\begin{figure*}[t]
    \centering
    \includegraphics[width=1.0\textwidth]{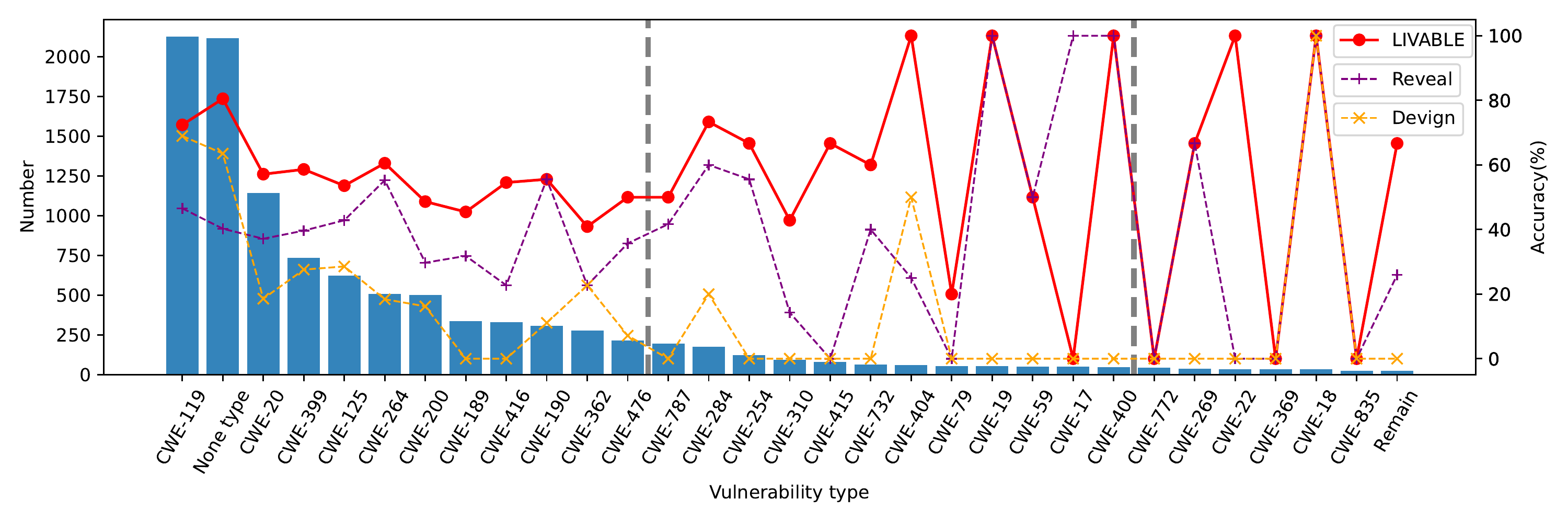}
    \caption{The accuracy of each vulnerability type in \tool. The X-axis denotes the vulnerability type. The Y-axis on the left and right indicate the number of vulnerabilities and accuracy respectively. The lines in yellow, purple, and red represent the accuracy of Devign, Reveal, and \tool respectively. }
    \label{cweid}
\end{figure*}

Figure~\ref{cweid} shows the results of \tool with Devign and Reveal in terms of vulnerability type classification, with respect to each vulnerability type. In the head class, \tool performs better in the detection of unknown types (i.e. None type) of vulnerabilities. This may be due to that the vulnerability representation learning module captures more information from the code snippets and obtains a more discriminative vulnerability representation. In the medium and tail classes, \tool also performs better than the previous method, especially for CWE-787, CWE-284 and CWE-25, CWE-22, etc. This is due to the fact that \tool gives higher weighted attention to the tail class samples.
In general, \tool outperforms baselines for classifying vulnerability types.

In addition, we note that the performance of Reveal on the tail data is not affected by the loss functions used. %In addition, the Reveal performs more consistently on the tail data. %This may be due to the way they use SMOTE's sampling method, but it also leads them to lose accuracy in identifying the head classes.
This may be due to the way that the learned vulnerability representations are constrained by over-smoothing and the model has difficulty learning %for 
the samples of tail class.

\subsubsection{Compared with re-weighting methods}
In this subsection, we experiment with different re-weighting methods to solve the long-tailed problem of vulnerability type classification.

\textbf{Cross-entropy Loss:} When using the CE-loss, our \tool is more accurate in identifying head classes than tail classes. CE-Loss encourages the whole model to be over-confident in the head classes for massive data. The experiment shows that the head class samples performed 9.14\% and 22.15\% better than the medium and tail classes, respectively, in terms of accuracy. 
% But Reveal focused more on the tail classes by the SMOTE approach to re-sampling.

\textbf{Smoothing strategies:}
Smoothing strategies include label smooth loss and label-aware smooth loss. They are another regularization technique that encourages the model to be less over-confident in the head classes. We use these methods to replace CE loss and they both improve the accuracy of the head class by 2.89\% in \tool. %Unlike CE loss, label smoothing calculates the loss of the soft value of the label. It is effective by enhancing the competitive with loss-correction under label noise~\cite{DBLP:conf/icml/LukasikBMK20/labels, DBLP:journals/corr/abs-2301-05456/noise} in the classification of vulnerability types. 
The use of the smoothing strategies %smooth strategy 
in other methods did not work, which is probably due to the poor vulnerability representation they captured.

\textbf{Class-balanced strategies:}
The class-balanced strategies consist of the class-balanced loss and class-balanced focal loss, which assigns different weights for classes and instances. They lead to an average increase in accuracy of 2.94\% and 15.39\% for the medium and tail classes respectively, but an average decrease of 2.98\% for the head classes. It shows that class-balanced strategies focus more on the tail data but will cause a decrease in the accuracy of the head classes.

\textbf{Adaptive re-weighting module:}
Compared with other re-weighting methods, the adaptive re-weighting module achieves the highest accuracy in all four metrics. 
Specifically, the adaptive re-weighting module achieves an absolute improvement of 4.24\% in overall accuracy.%a 4.24\% performance improvement in overall accuracy.  
%In the head, medium, and tail classes, it improves the average accuracy by 4.13\%, 5.15\% and 5.13\% respectively. 
The tail-focused branch %tail learning branch 
leverages the advantages of class-balanced strategies, which assign higher weights to the more challenging tail samples in the early stages of training. It improves the average accuracy in medium and tail classes by 5.15\% and 5.13\% respectively. In the later phase of training, the head-focused branch %head learning branch 
involves the smoothing strategy, which calculates the loss of the soft value of the label. It is effective as a means of coping with label noise ~\cite{DBLP:conf/icml/LukasikBMK20/labels, DBLP:journals/corr/abs-2301-05456/noise} in the vulnerability types classification. The shifting weight ensures that the head and tail classes can be constantly learned in the whole training process, which avoids conflicts with each other.
In summary, the adaptive re-weighting module is able to learn representations of tail data efficiently based on a limited number of samples.
% In the tail classes classification, our proposed method has the same accuracy as the class-balanced strategies, but in the head and medium classes, we improve the accuracy by 7.16\% and 4.41\%, respectively.

 \begin{tcolorbox}
 \textbf{Answer to RQ1:} 
\tool outperforms all the baseline methods in terms of accuracy for head classes, medium classes, and tail classes, as well as for all samples. %\tool outperforms all the baseline methods in head classes, medium classes, tail classes, and all sample accuracy in the vulnerable type classification. 
In particular, the adaptive re-weighting module achieves better performance than existing re-weighting methods.
 \end{tcolorbox}

\subsection{RQ2: Performance on Vulnerability Detection}
%RQ2: How effective is vulnerability representation learning module in vulnerability detection?

To answer this research question, we explore the performance of the vulnerability representation learning (VRL) module in \tool and compare it with other baseline methods. Table~\ref{e1_effective} shows the overall results.
%To answer this research question, we first explore the performance of \chen{\tool for} vulnerability detection and compare it with other baseline methods.
%Table~\ref{e1_effective} shows the overall results of all baseline methods and the vulnerability representation learning module performance on the three datasets.

\textbf{The proposed VRL module %methods 
outperform all the baseline methods:} Overall, VRL module %\tool 
achieves better results on the FFMPeg+Qemu, Reveal and Fan \et datasets. When considering all the performance metrics
regarding the three datasets (12 combination cases), VRL module %\tool 
obtains the best performance in 11 out of 12 cases. On all three datasets, VRL module %\tool 
outperforms all baseline methods in terms of accuracy and F1 score metrics. Especially in terms of F1 score, the relative improvements are 8.36\%, 23.74\%, and 68.78\%%63.7\%
, respectively. Compared to the best-performing baseline method Reveal, VRL module %\tool 
achieves an average improvement of 9.88\%, 67.49\%, 2.09\%, and 33.63\% on the four metrics across all datasets. This indicates that the vulnerability representation learning module performs better on vulnerability detection than previous methods, obtaining a more discriminative code representation.

\textbf{The combination of the graph-based branch and the token-based branch can achieve better performance:} The experimental results also show that the graph-based methods perform better in most cases. The token-based methods perform well on some metrics. For example, in Fan \et, the token-based methods obtain better performance than graph-based methods on precision metrics overall. The reason may be attributed to that token-based methods are better at capturing semantic information in the code, whereas graph-based methods are more attentive to the structure information in the code. Thus, VRL module %\tool 
combines the advantages of both graph-based and token-based methods to capture more information and obtain a more discriminating code representation in vulnerability detection.

\begin{table*}[h]
\centering

\setlength{\tabcolsep}{1.2mm}
\renewcommand{\arraystretch}{1.2}

\caption{
Comparison results between vulnerability representation learning
(VRL) module  %\tool 
and the baselines on the three datasets in vulnerability detection. ``-'' means that the baseline does not apply to the dataset in this scenario. The best result for each metric is highlighted in bold. The shaded cells represent the performance of the top-3 best methods in each metric. Darker cells represent better performance.}
\resizebox{.97\textwidth}{!}{
\begin{tabular}{l|cccc|cccc|cccc}
\toprule
\diagbox{Metrics(\%) }{Dataset} & \multicolumn{4}{c|}{FFMPeg+Qemu \cite{DBLP:conf/nips/ZhouLSD019/}}        & \multicolumn{4}{c|}{Reveal \cite{DBLP:journals/tse/ChakrabortyKDR22/}}             & \multicolumn{4}{c}{Fan \et. \cite{DBLP:conf/msr/FanL0N20/}}                \\
\midrule
Baseline                         & Accuracy & Precision & Recall & F1 score     & Accuracy & Precision & Recall & F1 score    & Accuracy & Precision & Recall & F1 score    \\
\midrule
VulDeePecker                    & 49.61   & 46.05    & 32.55 & 38.14 & 76.37   & 21.13    & 13.10 & 16.17 & 81.19   & \cellcolor{gray!45}38.44    & 12.75 & 19.15 \\

Russell \et.                  & \cellcolor{gray!20}57.60   & \cellcolor{gray!20}54.76    & 40.72 & 46.71 & 68.51   & 16.21    & \cellcolor{gray!20}52.68 & 24.79 & 86.85   & 14.86    & \cellcolor{gray!20}{26.97} & 19.17 \\

SySeVR                          & 47.85   & 46.06    & 58.81 & 51.66 & 74.33   & \cellcolor{gray!45}40.07  & 24.94 & 30.74   & \cellcolor{gray!20}90.10   & \cellcolor{gray!20}30.91   & 14.08 & 19.34 \\

Devign                          & 56.89   & 52.50    &  \cellcolor{gray!20}64.67 &  \cellcolor{gray!20}57.95 & \cellcolor{gray!45}87.49   & \cellcolor{gray!20}31.55    & 36.65 & \cellcolor{gray!20}33.91 & \cellcolor{gray!45}{92.78}   & 30.61   & 15.96   & \cellcolor{gray!20}20.98 \\

Reveal                          & \cellcolor{gray!45}61.07   & \cellcolor{gray!45} 55.50    & \cellcolor{gray!45} 70.70 & \cellcolor{gray!45}62.19 &  \cellcolor{gray!20}81.77   &  \cellcolor{gray!20}31.55    & \cellcolor{gray!70}\textbf{61.14}& \cellcolor{gray!45}41.62   & 87.14   & 17.22   & \cellcolor{gray!45}34.04 & \cellcolor{gray!45}22.87 \\

IVDetect                        & 57.26   & 52.37    & 57.55 & 54.84 & -   & -         & -      & -      & -   & -         & -      & -      \\
\midrule
%Our proposed         
VRL module
& \textbf{\cellcolor{gray!70}64.84}   & \cellcolor{gray!70}\textbf{57.87}    & \cellcolor{gray!70}\textbf{80.67} &\cellcolor{gray!70} \textbf{67.39} &\cellcolor{gray!70}\textbf{93.53}   & \cellcolor{gray!70} \textbf{52.27}    &  \cellcolor{gray!45}50.55 & \cellcolor{gray!70}\textbf{51.50} & \cellcolor{gray!70}\textbf{95.05}   & \cellcolor{gray!70}\textbf{40.04}  & \cellcolor{gray!70}\textbf{37.27} & \cellcolor{gray!70}\textbf{38.60}\\
\bottomrule
\end{tabular}}

\label{e1_effective}
\end{table*}

 \begin{tcolorbox}
 \textbf{Answer to RQ2:} 
The vulnerability representation learning module outperforms all baseline methods in terms of accuracy and F1 scores in vulnerability detection.
 In particular, it improves the accuracy of the three datasets by 3.77\%, 6.04\% and 2.27\% respectively over the best baseline method.
 \end{tcolorbox}

 \subsection{{RQ3: Ablation Study}
 %RQ3: What is the impact of different components on the type classification performance of \tool?
 }
To answer this research question, we explore the effect of %differentiated propagation-based GNN, sequence-to-sequence model,
vulnerability representation learning
module and adaptive re-weighting module on the performance of \tool by performing an ablation study on vulnerability type classification.
%To answer this research question, we first explore the effect of graph and sequence \chen{representation } for \tool. Then, we explore the effect of the adaptive re-weighting module on the performance of \tool by performing an ablation study on vulnerability type classification. 
\subsubsection{Vulnerability representation learning module}
To explore the contribution of the vulnerability representation learning module, we construct the following two variants in the module for comparison: 
%In this section, we first explore the contribution of the vulnerability representation learning module to the performance. We construct the following two branch variations in the backbone for comparison:
(1) only using differentiated propagation-based GNN (denoted as w/o sequence representation) to validate the impact of the sequence-to-sequence model (2) only using sequence-to-sequence model (denoted as w/o graph representation) to verify the effectiveness of differentiated propagation-based GNN.
%(1) only using \chen{differentiated propagation-based GNN }(denoted as w/o \chen{sequence-to-sequence model}):   we remove the sequence \chen{representation } to validate the impact of the sequence \chen{representation} in vulnerability type classification; (2) only use the sequence \chen{representation } (denoted as w/o graph \chen{representation }): we remove the graph \chen{representation } to verify the effectiveness of the graph \chen{representation};

As shown in Table~\ref{e3_ablation}, all variations contribute positively to the overall performance. This indicates that the combination of differentiated propagation-based GNN %graph representation 
and 
sequence-to-sequence model %sequence representation 
can better help the model to capture more discriminative representations from the source code. Overall, the impact of combining sequence-to-sequence model %sequence representation 
is greater, which achieves a 14.65\% improvement while combining differentiated propagation-based GNN %graph representation 
gives only a 5.41\% performance improvement.

\subsubsection{Adaptive re-weighting module}
Then we explore the contribution of the adaptive re-weighting module to the performance of vulnerable type classification.
We construct three variations, including (1) using the cross-entropy loss instead of the adaptive re-weighting module (denoted as w/o adaptive) (2) removing the tail-focused branch while keeping the head-focused branch (denoted as w/o tail-focused branch) (3) removing head-focused branch while keeping tail-focused branch (denoted as w/o head-focused branch).%the tail classes learning the branch is removed while keeping the head classes learning branch (denoted as w/o tail branch) (3) the head classes learning branch is removed while keeping the tail branch (denoted as w/o head branch).

The results of ablation studies are shown in Table~\ref{e3_ablation}. We find that the performance of all the variants is lower than \tool, which indicates that all the components contribute to the overall performance of \tool. The head-focused branch %head branch 
has more influence on the differentiated propagation-based GNN%graph \chen{representation} %branch
, while the tail-focused branch %tail branch 
enables the sequence-to-sequence model %sequence \chen{representation }%branch 
to have a greater performance improvement.
   
\begin{table*}[h]
\centering

\setlength{\tabcolsep}{1.2mm}
\renewcommand{\arraystretch}{1.2}

\caption{Results of ablation study in vulnerability type classification.}
\resizebox{.97\textwidth}{!}{
\begin{tabular}{l|rrrr|rrrr|rrrr}
\toprule
\diagbox{Metrics(\%) }{Methods} & \multicolumn{4}{c|}{w/o sequence representation}        & \multicolumn{4}{c|}{w/o graph representation}             & \multicolumn{4}{c}{\tool}                \\
\midrule
Methods                & \multicolumn{1}{c}{Head}    & \multicolumn{1}{c}{Medium}  & Tail    & Accuracy  & Head    & Medium  & Tail    & Accuracy  & Head    & Medium  & Tail    & Accuracy       \\
\midrule
w/o adaptive         & 46.95 & 25.00 & 38.46 & 44.67 & 56.59 & 32.35 & 38.46 & 53.91 & 60.61 & 51.47 & 38.46 & 59.32 \\
w/o tail-focused branch & 49.20 & 44.12 & 30.77 & 48.36 & 60.29 & 47.06 & 38.46 & 58.61 & 63.50 & 55.88 & 46.15 & 62.45 \\
w/o head-focused branch      & 48.71 & 33.82 & 30.77 & 46.94 & 62.38 & 51.47 & 46.15 & 61.02 & 61.41 & 45.59 & 38.46 & 59.46 \\
\midrule
%Our proposed         
\tool
                      & -   & -         & -      & -      & -   & -         & -      & -               & \textbf{64.79}& \textbf{58.82}& \textbf{53.85}& \textbf{64.01}     \\
\bottomrule 
\end{tabular}}
\label{e3_ablation}
\end{table*}

 \begin{tcolorbox}
 \textbf{Answer to RQ3:} 
The vulnerability representation learning module combines the differentiated propagation-based GNN and sequence-to-sequence model, in which both %sequence \chen{representation }%branch 
contributes significantly to the performance of \tool, with an improvement of 5.41\% and 14.65\% respectively in terms of accuracy. The adaptive re-weighting module also has a positive effect on model performance.
 \end{tcolorbox}
 
 \subsection{RQ4: Parameter Analysis}
 %RQ4:  What is the influence of hyper-parameters on the performance of \tool ?
 
  \begin{table}[t]
\centering
\setlength{\tabcolsep}{2mm}

\renewcommand{\arraystretch}{1.1}

\caption{The impact of the number of GNN layers in the graph branch and hidden size in the sequence branch on the performance of \tool. 
}
\begin{tabular}{c|c|c|c}
\toprule
\multicolumn{2}{c|}{Graph Branch} & \multicolumn{2}{c}{Sequence Branch} \\
\midrule
Layer number    & Accuracy(\%)   & Hidden size      & Accuracy(\%)    \\
\midrule
12                  & 62.07    & 128                 & 56.53       \\
14                  & 62.78    & 256                 & 59.23       \\
\textbf{16}                  & \textbf{64.01}    & \textbf{512}                 & \textbf{64.01}       \\
18                  & 62.92    & 768                 & 62.07       \\
20                  & 61.50    & 1024                & 62.93       \\
\bottomrule
\end{tabular}
\label{hyper}
\end{table}

To answer this research question, we explore the impact of hyper-parameters in the \tool, including the layer number of the differentiated propagation-based GNN% branch
, and the hidden size in the sequence-to-sequence model.%sequence branch.

\subsubsection{Number of GNN layers}
 
Table~\ref{hyper} shows the accuracy of \tool with different layers of GNN on the vulnerability type classification. In the previous work, GNNs usually achieve the best performance at 2 layers. However, with the number of layers set to 16, \tool achieves a best performance of 64.01\%. This indicates that \tool can learn the relationships between more distant nodes than traditional methods, effectively alleviating the problem of over-smoothing. In addition, due to the limited number of nodes in the code structure graph, the number of layers of the GNN model cannot be increased indefinitely, and the accuracy rate starts to decrease when the number of layers equals 18. In summary, the model benefits from the use of more layers to capture relationships between distant nodes, enhancing the model's ability to capture information about the graph structure.

\subsubsection{Size of hidden layers}
We also experiment with MLPs with different sizes of hidden neurons. We use the same setup in the previous experiments. We experiment with the task of vulnerability type classification. The results are shown in Table~\ref{hyper} and show that the \tool performs best when using a hidden size of 512. This may be due to the fact that the task requires the detection of a larger number of vulnerability types. When more dimensions are used (more than 512), the performance starts to drop, which may be due to the over-fitting problem.%more complex model.

 \begin{tcolorbox}
 \textbf{Answer to RQ4:} 
The hyper-parameter settings of GNN layer number and hidden size can impact the performance of \tool. The proposed differentiated propagation-based GNN %branch 
can use more layers for learning to distinguish node representations.
 \end{tcolorbox}

\section{Discussion}
\label{sec:discussion}
%\subsection{Why does \tool alleviate over-smoothing?}

\subsection{Why Does \tool Work?}

\begin{figure*}[t]
	\centering
    \includegraphics[width=1.0\textwidth]{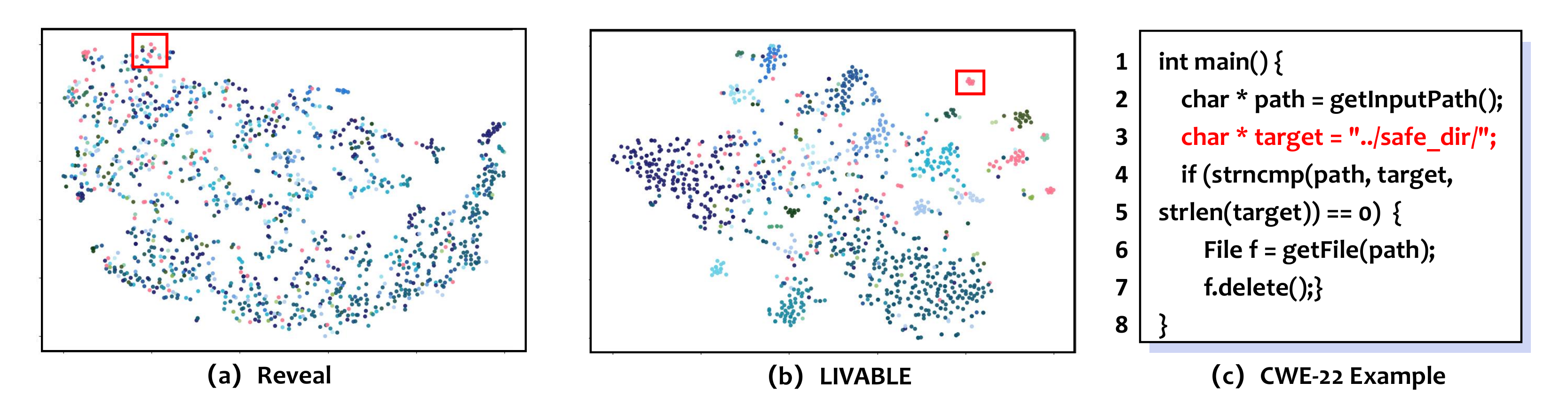}
	\caption{The case study of \tool, which mainly contains three parts: (a) T-SNE visualization of Reveal, (b) T-SNE visualization of \tool , and (c) a CWE-22~\cite{CWEID22} example in tail class. In (a) and (b), the blue dots represent head samples, the green dots represent mid samples and the pink dots represent tail samples. In (c), the red-colored code is vulnerable code. }
	\label{case}
\end{figure*}

In this section, we identify the following two advantages of \tool, which
can explain its effectiveness in vulnerability type classification.
% detection.
%To understand whether \tool pays attention to the medium and tail classes in the dataset, 
% And 
We visualize the vulnerability representations learnt by \tool and the best baseline Reveal via the popular T-SNE technique~\cite{van2008visualizing/TSNE}, as shown in Figure~\ref{case} (a) and (b), respectively.
% As shown in Figure~\ref{case} (a) and (b), we compare the Reveal and \tool. We also provide a theoretical analysis to show that \tool alleviates the problem of over-smoothing.

% \begin{equation}
% \label{propagation}
% H^{l} = \frac{1}{l} \sum_{i=0}^{l-1} \left ( \left ( 1-\alpha \right ) \left ( {\widetilde{D}^{-\frac{1}{2}}}{\widetilde{A}}{\widetilde{D}^{-\frac{1}{2}}}\right )^{i} H^{0} \right )+ \alpha H^{(0)}  
% \end{equation}

\textbf{(1) The ability to learn the vulnerability representation.} 
The proposed vulnerability representation learning module helps \tool to learn the vulnerability representation. More specifically, the proposed differentiated propagation GNN branch alleviates the over-smoothing problem and the sequence branch enhances %enhanced 
the model's ability for capturing semantics information. As shown in Figure~\ref{case} (a) and (b), compared with the Reveal, \tool can enhance the discriminability of representation on the type classification, in which head and tail classes are clustered more tightly.
% But some clusters are mixed tail classes together\gsz{[this sentence can be removed?]}.
%CE loss focuses more on the identification of the head class, which leads to over-confidence in the head class. Furthermore, some of the tail classes are attached to a larger number of classes, and the distance between classes is too small. For example, the pink sample in the red box is CWE-22, and the excessive proximity between classes with the green sample causes it to be completely unrecognized (accuracy is 0).

\textbf{(2) The ability to capture the representation of the tail classes.}
Although Reveal also identifies %identified 
some samples from the tail class, they are %were 
usually mixed and clustered with samples from other classes.  
In comparison, \tool is easier to detect samples of the tail class. We choose CWE-22 in tail classes as an example, as shown in Figure~\ref{case} (c). The example code attempts to validate a given input path by checking it against an allowlist and once validated delete the given file. But the ``../" sequence in line 3 will cause the program to delete the important file in the parent directory. As can be seen in Figure~\ref{case} (b), the CWE-22 sample in the red box can be identified because its
% the 
distance with
% between it and 
the other class samples are clearer.

\subsection{Threats to Validity}
One threat to validity comes from the size of our constructed dataset. %Another threat is the dataset. 
Following the previous methods, we use a C/C++ dataset built for vulnerability detection. In the task of vulnerability type classification, we extract 91 types of vulnerability functions from this dataset and construct the new dataset. However, in the real world, there are more than 300 types of vulnerabilities. %Therefore, our \tool is not necessarily applicable to other data. 
%To mitigate this threat, the model used in \tool can be easily trained, which makes it more suitable to be employed in real-world long-tailed distribution scenarios with larger amounts of data. 
In the future, we will collect a more realistic benchmark for evaluation.

The second threat to validity is that our proposed \tool has only experimented on the C/C++ dataset. Experiments have not been conducted on datasets from other programming languages, such as Java and Python. In the future, we will select more programming languages and corresponding datasets to further evaluate the effectiveness of \tool.
%our method's effectiveness.

%\yun{More theat?}
Another threat comes from the implantation of baselines. Since Devign~\cite{DBLP:conf/nips/ZhouLSD019/} does not publish their implementation and hyper-parameters. So we reproduce Devign based on the
Reveal’s~\cite{DBLP:journals/tse/ChakrabortyKDR22/} implementation and re-implement the method to the best of our abilities.

\section{Related Work}
\label{sec:related}

\subsection{Learning-Based Vulnerability Detection}
%\emph{B. Machine-Learning-Based Vulnerability Detection}
In recent years, learning-based methods have been widely used for vulnerability detection tasks. The researcher first uses the Machine Learning (ML)-based method~\cite{DBLP:conf/sp/ForrestHSL96/, DBLP:conf/uss/YamaguchiLR11a/, DBLP:conf/softcomp/SantosDBNB12/, DBLP:conf/ccs/NeuhausZHZ07/, DBLP:journals/tse/ShinMWO11/} in vulnerability detection. For example, Neuhaus\et~\cite{DBLP:conf/usenix/NeuhausZ09/} extract the dependency information matrix and vulnerability vectors from the source code, and use the support vector machine to detect them. %\lf{Grieco \et~\cite{DBLP:conf/codaspy/GriecoGURFM16/} use the features which abstract the use patterns of the C standard library in several machine learning models (i.e., logistic regression, MLP of single hidden layer, random forest).} 
Grieco\et~\cite{DBLP:conf/codaspy/GriecoGURFM16/} abstract features from the C-standard library and use multiple machine learning models, such as logistic regression and random forest.

ML-based methods rely heavily on manual feature extraction, which is time-consuming and may require much effort. Therefore, Deep Learning (DL)-based methods~\cite{DBLP:conf/ccs/LinZLPX17/, DBLP:conf/qrs/LiHZL17/, DBLP:conf/nips/HarerOLRRKC18/, DBLP:conf/icse/WangLT16/, DBLP:conf/kbse/WhiteTVP16/} learn the input representation from the source code, which can better capture the vulnerability patterns.
%In recent years, a great number of learning-based vulnerability detection methods have demonstrated exceptional performance. %Neuhaus \et~\cite{DBLP:conf/usenix/NeuhausZ09/} use support vector machines to detect vulnerabilities. Grieco \et~\cite{DBLP:conf/codaspy/GriecoGURFM16/} add lightweight static and dynamic features in the machine learning model to detect vulnerabilities.
Depending on the input generated from source code and training model types,  DL-based approaches can group into two different types: sequence-based and graph-based methods. 
Sequence-based approaches utilize source code tokens as their model inputs. 
 %Russell \et~\cite{DBLP:conf/icmla/RussellKHLHOEM18/} build the model on CNN.
 VulDeePecker~\cite{DBLP:conf/ndss/LiZXO0WDZ18/} extracts data flow information from source code and adapts BiLSTM to detect buffer error vulnerabilities.
 %Russell \et~\cite{DBLP:conf/icmla/RussellKHLHOEM18/} divides each code fragment at the function level and treats them into a CNN model.
 %DeepBugs~\cite{DBLP:journals/pacmpl/PradelS18/} is a framework for learning-based and name-based bug detection which transfers a program to a vector.
 SySeVR~\cite{DBLP:journals/tdsc/0027ZX0ZC22/} combines both control flow and data flow to generate program slices and utilizes a bidirectional RNN for code vulnerability detection.  %using RNN to detect whether a program has various types of vulnerabilities. 
 Graph-based methods capture more structural information than sequence-based methods, which achieve better performance on vulnerability detection tasks. %In view of this, graph-based methods have emerged as a viable alternative. 
 %VGDetector~\cite{DBLP:conf/iceccs/ChengWHZXYS19/} constructs a control-flow graph from source code, and adapts them into the GCN model to detect control-flow-related vulnerabilities.
 Devign~\cite{DBLP:conf/nips/ZhouLSD019/} builds graphs combining AST, CFG, DDG and NCS edges and uses the GGNN model for vulnerability detection. %Reveal~\cite{DBLP:journals/tse/ChakrabortyKDR22/} uses a combination of GGNN, MLP and Triplet Loss to construct the model, where the code property graph is used as input representation. 
 IVDetect~\cite{DBLP:conf/sigsoft/Li0N21/} constructs the Program Dependency Graph
(PDG) and proposes feature attention GCN to learn the graph representation. %VELVET~\cite{DBLP:conf/wcre/DingSZLMKR22/}, a novel ensemble learning approach to locate vulnerable statements based on the scores learned from GGNN and transformer~\cite{transformer}.
 %\lf{The token-based methods can capture global interactions and temporal information in code snippets. But in representing the structural information, token-based methods demonstrate poor performance. Thus, we propose to combine token-based and graph-based methods.}
 
However, directly adopting these approaches tends to result in poor performance in vulnerability type classification, which is limited by class-imbalanced and over-smoothing problems in GNNs. In this paper, we propose a vulnerability representation learning module to boost the model to learn the vulnerability representation and capture vulnerability types.

\subsection{Class Re-weighting Strategies}
Long-tailed classification has attracted increasing attention due to the prevalence of imbalanced data in real-world applications~\cite{DBLP:conf/cvpr/ZhouCWC20/bbn, DBLP:conf/cvpr/SzegedyVISW16/, DBLP:journals/corr/abs-2110-04596/longtail, DBLP:conf/cvpr/TanWLLOYY20/reweight1,DBLP:conf/iccv/ParkLJ021/reweight3}. 
It leads to a small portion of classes having massive sample points but the others contain only a few samples, which makes the model ignore the identification of tail classes.
Recent studies have mainly pursued re-weighting strategies.
For example, Lin \et propose Focal Loss~\cite{DBLP:conf/iccv/LinGGHD17/}, which adjusts the standard CE-loss to reduce the relative loss for well-classified samples and focus more on rare samples that are misclassified during model training. Cui \et~\cite{DBLP:conf/cvpr/CuiJLSB19/} design the class-balanced loss, which adds a weight related to the number of samples in the class to make the model focus on the tail class. Another effective approach is to reduce the model's excessive focus on head classes. Zhong \et~\cite{DBLP:conf/cvpr/ZhongC0J21/} use a label smoothing strategy and combines the probability distributions of different classes to mitigate the overweighting of the head classes.
In this paper, we propose an adaptive re-weighting module to learn the vulnerability representation in the tail classes.

\section{Conclusion}
\label{sec:conclusion}
% In this paper, we systematically study data distribution of vulnerability type classification to effectively pin down the real world vulnerability type. 
% We empirically show the distribution of vulnerability types in the real world belongs to the long-tail distribution and the importance of the tail class vulnerabilities that the neglect of the tail class may pose a huge security risk in the real world.
% Our investigation found that existing methods are hard to learn real world vulnerability types in the tail classes. And existing techniques do not consider data imbalance and long-tailed distribution in vulnerability prediction. 
% Following these empirical findings, we propose a framework, \tool, for considering the tail class vulnerabilities in the real world. It involves (1) a two-branch backbone to combine the graph branch and sequence branch to learn more discriminative code representation, and (2) a temperature loss to learn both head-classes learning and tail-classes representations simultaneously. In summary, this paper clarifies the potential problems with existing DL-based vulnerability type classification systems and demonstrates \tool potential towards a better vulnerability prediction tool.
In this paper, we show the distribution of vulnerability types in the real world belongs to the long-tailed distribution and the importance of the tail class vulnerabilities.
We propose \tool, a long-tailed software vulnerability type classification approach, which involves the vulnerability representation learning module and adaptive re-weighting module. It can distinguish node representation and enhance vulnerability representations. It can also predict vulnerability types in the long-tailed distribution. Our experimental results on vulnerability type classification validate the effectiveness of \tool, and the results in vulnerability detection and the ablation studies further demonstrate the advantages of \tool. The future works include combining the vulnerability repair with type and generating human-readable or explainable reports due to the vulnerability type.

\section*{Data availability}
%\textbf{Data availability}: 
Our source code as well as experimental data are available at: \textit{{\http}}.
\bibliographystyle{IEEEtran}
\bibliography{IEEEabrv, sample-base}
\end{document}